\documentstyle[12pt]{article}
\newlength{\extraspace}
\newlength{\extraspaces}
\newcommand{\be}{\begin{equation}
\addtolength{\abovedisplayskip}{\extraspaces}
\addtolength{\belowdisplayskip}{\extraspaces}
\addtolength{\abovedisplayshortskip}{\extraspace}
\addtolength{\belowdisplayshortskip}{\extraspace}}
\newcommand{\ee}{\end{equation}}
\newcommand{\ba}{\begin{eqnarray}
\addtolength{\abovedisplayskip}{\extraspaces}
\addtolength{\belowdisplayskip}{\extraspaces}
\addtolength{\abovedisplayshortskip}{\extraspace}
\addtolength{\belowdisplayshortskip}{\extraspace}}
\newcommand{\ea}{\end{eqnarray}}
\newcommand{\nonu}{\nonumber \\[.5mm]}
\newcommand{\A}{&\!\!\!}
\begin{document}
\thispagestyle{empty}
\setlength{\baselineskip}{6mm}
%
%
\vspace*{7mm}
\begin{center}
{\large\bf Commutator-based linearization of \\[2mm]
$N = 1$ nonlinear supersymmetry
} \\[20mm]
%
%
{\sc Motomu Tsuda}
\footnote{
\tt e-mail: tsuda@sit.ac.jp} 
\\[5mm]
{\it Laboratory of Physics, 
Saitama Institute of Technology \\
Fukaya, Saitama 369-0293, Japan} \\[20mm]
\begin{abstract}
We consider the linearization of $N = 1$ nonlinear supersymmetry (NLSUSY) 
based on a commutator algebra in Volkov-Akulov NLSUSY theory. 
We show explicitly that $U(1)$ gauge and scalar supermultiplets in addition to a vector supermultiplet 
with general auxiliary fields in linear SUSY theories are obtained from 
a same set of bosonic and fermionic functionals (composites) which are expressed as simple products of 
the powers of a Nambu-Goldstone fermion and a fundamental determinant in the NLSUSY theory. 
\\[5mm]
%
%
\noindent
PACS: 11.30.Pb, 12.60.Jv, 12.60.Rc \\[2mm]
\noindent
Keywords: supersymmetry, Nambu-Goldstone fermion, commutator algebra, supermultiplet 
\end{abstract}
\end{center}

\newpage

\noindent
Linear supersymmetric (LSUSY) theories \cite{WZ,WB} and Volkov-Alulov (VA) nonlinear SUSY (NLSUSY) one \cite{VA} 
are related to each other as shown explicitly for $N = 1$ and $N = 2$ SUSY \cite{IK}-\cite{STT2}. 
In these relations (NL/LSUSY relations), component fields for linear supermultilpets are expressed 
as functionals (composites) of Nambu-Goldstone (NG) fermions in the VA NLSUSY theory, 
which give LSUSY transformations of the component fields under NLSUSY ones of the NG fermions 
and lead to the relations between VA NLSUSY and LSUSY actions. 

The NL/LSUSY relations are obtained systematically in superspace formalism 
by means of superfields on specific superspace coordinates depending on the NG fermions \cite{IK} 
(for recent review, see \cite{Iv}). 
On the other hand, in the component expression, we have recently discussed a linearization procedure of NLSUSY 
based on a commutator algebra in the VA NLSUSY theory for extended SUSY \cite{MT}: 
In the VA NLSUSY theory, the NLSUSY action describing dynamics of NG fermions is defined from a fundamental determinant 
which corresponds to both a spontaneous SUSY breaking and a basic geometrical structure of spacetime with Grassmann coordinates. 
We have introduced a set of functionals which are expressed in terms of products of NG fermions 
and the fundamental determinant in the NLSUSY theory and shown that general LSUSY transformations 
of basic components defined from the set of those functionals are uniquely determined 
based on the fact that every functional of NG fermions and their derivative terms 
satisfies the commutation relation in the NLSUSY theory under NLSUSY transformations. 

In addition to the superspace formalism for the NL/LSUSY relations, 
the above commutator-based linearization procedure of NLSUSY would be also useful, 
e.g. in order to understand low-energy physics in NLSUSY general relativistic (GR) theory \cite{KS,ST1}. 
In the NLSUSY-GR theory, an Einstein-Hilbert-type fundamental action is defined from the NG fermions and a vierbein. 
It possesses symmetries which are isomorphic to $SO(N)$ super-Poincar\'e group 
and contains the VA NLSUSY action in the cosmological term, which express a dimensional constant in the NLSUSY theory 
in terms of the cosmological and gravitational constant. 
Therefore, it is important for the NLSUSY-GR theory to know the NL/LSUSY relations 
for $N$-extended SUSY in more detail, including up to (all-order) functional (composite) expressions of the NG fermions 
for component fields in LSUSY theories. 

In this letter, towards the basic understanding of the linearization of NLSUSY for extended SUSY in the component expression, 
the $N = 1$ NL/LSUSY relations are reconsidered by applying the linearization procedure based on the commutator algebra, 
though the basic relations between LSUSY actions (with a $D$-term) and the VA NLSUSY one for $N = 1$ SUSY was established 
in terms of superfields and SUSY invariant constraints \cite{IK,Ro,STT1}. 
By introducing bosonic and fermionic functionals which are expressed as simple products of the powers of a NG fermion 
(with gamma matrices) and the fundamental determinant in the VA NLSUSY theory, 
we linearize $N = 1$ NLSUSY and construct LSUSY transformations of basic components defined from the set of the functionals, 
which just corresponds to LSUSY transformations for the vector supermultiplet with general auxiliary fields \cite{WZ,WB}. 
We also show that both $U(1)$ gauge and scalar supermultiplets are derived from 
appropriate recombinations of the functionals for the basic components in the LSUSY transformations, 
so that the common NG-fermion-functional (composite) structure of component fields 
in the linear supermultiplets is manifest. 

Let us briefly review the VA NLSUSY theory \cite{VA} and properties of functionals of the NG fermions \cite{MT} for extended SUSY. 
For $N$-Majorana NG fermions $\psi^i$, their NLSUSY transformations 
are defined by means of constant (Majorana) spinor parameters $\zeta^i$ as 
\footnote{
The indices $i,j, \cdots = 1, 2, \cdots, N$ 
and Minkowski spacetime indices are denoted by $a, b, \cdots = 0, 1, 2, 3$. 
Gamma matrices satisfy $\{ \gamma^a, \gamma^b \} = 2 \eta^{ab}$ 
with the Minkowski spacetime metric $\eta^{ab} = {\rm diag}(+,-,-,-)$ 
and $\displaystyle{\sigma^{ab} = {i \over 4}[\gamma^a, \gamma^b]}$ is defined. 
}
\be
\delta_\zeta \psi^i = {1 \over \kappa} \zeta^i + \xi^a \partial_a \psi^i, 
\label{NLSUSY}
\ee
where $\xi^a = i \kappa \bar\psi^j \gamma^a \zeta^j$. 
The NLSUSY transformations (\ref{NLSUSY}) satisfy a commutator algebra, 
\be
[\delta_{\zeta_1}, \delta_{\zeta_2}] = \delta_P(\Xi^a), 
\label{NLSUSYcomm}
\ee
where $\delta_P(\Xi^a)$ means a translation with parameters $\Xi^a = 2 i \bar\zeta_1^i \gamma^a \zeta_2^i$. 
Under the NLSUSY transformations (\ref{NLSUSY}), a Lorentz tensor which is defined by means of 
$w^a{}_b = \delta^a_b + t^a{}_b$ with $t^a{}_b = - i \kappa^2 \bar\psi \gamma^a \partial_b \psi$ 
and the determinant $\vert w \vert = \det w^a{}_b$ transform respectively as 
\ba
\A \A 
\delta_\zeta w^a{}_b = \xi^c \partial_c w^a{}_b + \partial_b \xi^c w^a{}_c, \\
\A \A 
\delta_\zeta \vert w \vert = \partial_a (\xi^a \vert w \vert). 
\label{detw}
\ea
The VA NLSUSY action is defined based on the fundamental determinant $\vert w \vert$; 
\be
S_{\rm NLSUSY} = - {1 \over {2 \kappa^2}} \int d^4x \ \vert w \vert, 
\label{NLSUSYaction}
\ee
where $\kappa$ is a dimensional constant whose dimension is (mass)$^{-2}$. 

In the NLSUSY theory, every Lorentz-tensor (or scalar) functional of the NG fermions and their derivative terms 
satisfies the commutator algebra (\ref{NLSUSYcomm}) under the NLSUSY transformations (\ref{NLSUSY}). 
That is, bosonic or fermionic functionals of $\psi^i$ and their derivative terms 
($\partial \psi^i$, $\partial^2 \psi^i$, $\cdots$, $\partial^n \psi^i$) with $\gamma$-matrices, 
which are denoted as 
\be
F^I_A = F^I_A(\psi^i, \partial \psi^i, \partial^2 \psi^i, \cdots, \partial^n \psi^i), 
\label{functionals}
\ee
satisfy the same commutation relation as Eq.(\ref{NLSUSYcomm}); 
\be
[\delta_{\zeta_1}, \delta_{\zeta_2}] F^I_A = \Xi^a \partial_a F^I_A, 
\label{NLSUSYcomm2}
\ee
where the indices $A$ means the Lorentz ones, e.g. $(a, ab, \cdots, {\rm etc.})$, 
and $I$ is the internal ones, e.g. $(i, ij, \cdots, {\rm etc.})$. 
The commutation relation (\ref{NLSUSYcomm2}) is proved from Eq.(\ref{NLSUSYcomm}) and the fact that 
the derivative terms ($\partial \psi^i$, $\partial^2 \psi^i$, $\cdots$, $\partial^n \psi^i$) 
and products of two kinds of the fuctionals $F^I_A$ and $G^J_B$ which are respectively defined 
as Eqs.(\ref{functionals}) and (\ref{NLSUSYcomm2}) satisfy the same commutation relation. 

Here we introduce bosonic and fermionic functionals which are expressed as simple products of the powers of $\psi^i$ 
and the fundamental determinant $\vert w \vert$. Namely, we define the following functionals, 
\ba
b^i{}_A{}^{jk}{}_B{}^{l \cdots m}{}_C{}^n \left( (\psi^i)^{2(n-1)} \vert w \vert \right) 
\A = \A \kappa^{2n-3} \bar\psi^i \gamma_A \psi^j \bar\psi^k \gamma_B \psi^l \cdots \bar\psi^m \gamma_C \psi^n \vert w \vert, 
\label{bosonic} 
\\
f^{ij}{}_A{}^{kl}{}_B{}^{m \cdots n}{}_C{}^p \left( (\psi^i)^{2n-1} \vert w \vert \right) 
\A = \A \kappa^{2(n-1)} \psi^i \bar\psi^j \gamma_A \psi^k \bar\psi^l \gamma_B \psi^m \cdots \bar\psi^n \gamma_C \psi^p \vert w \vert, 
\label{fermionic}
\ea
which mean 
\ba
\A \A 
b = \kappa^{-1} \vert w \vert, \ b^i{}_A{}^j = \kappa \bar\psi^i \gamma_A \psi^j \vert w \vert, 
\ b^i{}_A{}^{jk}{}_B{}^l = \kappa^3 \bar\psi^i \gamma_A \psi^j \bar\psi^k \gamma_B \psi^l \vert w \vert \cdots, 
\label{bosonic0}
\\
\A \A 
f^i = \psi^i \vert w \vert, \ f^{ij}{}_A{}^k = \kappa^2 \psi^i \bar\psi^j \gamma_A \psi^k \vert w \vert, 
\label{fermionic0}
\cdots, 
\ea
for $n = 1, 2, \cdots$, respectively. 
In these functionals, (Lorentz) indices $A, B, \cdots$ are used as ones for a basis of $\gamma$ matrices, 
$\gamma_A = {\bf 1}, i \gamma_5, i \gamma_a, \gamma_5 \gamma_a$ or $\sqrt{2} i \sigma_{ab}$. 
Note that $f^i$ in Eq.(\ref{fermionic0}) give the leading order of superchages $Q^i$ 
and vector components are included in the functionals $b^i{}_A{}^j$ in Eq.(\ref{bosonic0}) for $N \ge 2$ SUSY 
(because $\bar\psi \gamma_a \psi = 0$ for $N = 1$ SUSY). 
The definitions of the functionals (\ref{bosonic}) and (\ref{fermionic}) (or (\ref{bosonic0}) and (\ref{fermionic0})) 
terminate with $n = 2N + 1$ and $n = 2N$, respectively, because $(\psi^i)^n = 0$ for $n \ge 4N+1$. 

The variations of the functionals (\ref{bosonic}) and (\ref{fermionic}) 
under the NLSUSY transformations (\ref{NLSUSY}) and (\ref{detw}) of $\psi^i$ and $\vert w \vert$ 
indicate that the bosonic and fermionic functionals in Eqs.(\ref{bosonic}) and (\ref{fermionic}) 
are linearly transformed to each other: 
\ba
\delta_\zeta b^i{}_A{}^{jk}{}_B{}^{l \cdots m}{}_C{}^n 
\A = \A 
\kappa^{2(n-1)} \left[ \left\{ \left( \bar\zeta^i \gamma_A \psi^j + \bar\psi^i \gamma_A \zeta^j \right) 
\bar\psi^k \gamma_B \psi^l \cdots \bar\psi^m \gamma_C \psi^n + \cdots \right\} \vert w \vert \right. 
\nonu
\A \A 
\left. + \kappa \partial_a \left( \xi^a \bar\psi^i \gamma_A \psi^j \bar\psi^k \gamma_B \psi^l 
\cdots \bar\psi^m \gamma_C \psi^n \vert w \vert \right) \right], 
\label{variation1}
\\[1mm]
\delta_\zeta f^{ij}{}_A{}^{kl}{}_B{}^{ml \cdots n}{}_C{}^p
\A = \A 
\kappa^{2n-1} \left[ \left\{ \zeta^i \bar\psi^j \gamma_A \psi^k \bar\psi^l \gamma_B \psi^m \cdots \bar\psi^n \gamma_C \psi^p 
\right. \right. 
\nonu
\A \A 
\left. + \psi^i \left( \bar\zeta^j \gamma_A \psi^k + \bar\psi^j \gamma_A \zeta^k \right) 
\bar\psi^l \gamma_B \psi^m \cdots \bar\psi^n \gamma_C \psi^p + \cdots \right\} \vert w \vert 
\nonu
\A \A 
\left. + \kappa \partial_a \left( \xi^a \psi^i \bar\psi^j \gamma_A \psi^k \bar\psi^l \gamma_B \psi^m 
\cdots \bar\psi^n \gamma_C \psi^p \vert w \vert \right) \right]. 
\label{variation2}
\ea
In fact, the functionals (\ref{bosonic}) and (\ref{fermionic}) lead to LSUSY transformations of component fields 
for (massless) vector linear supermultiplets with general auxiliary fields prior to transforming to gauge supermultiplets, 
e.g. in the case for $N = 2$ SUSY in two-dimensional spacetime \cite{ST3,ST4}. 
In addition, we have recently shown that LSUSY transformations for basic components 
which are defined from the set of the functionals (\ref{bosonic}) and (\ref{fermionic}) are uniquely determined 
from the variations (\ref{variation1}) and (\ref{variation2}) 
by examining the commutation relation (\ref{NLSUSYcomm2}) on the functionals \cite{MT}. 

In the following part of this letter, we use the functionals (\ref{bosonic}) and (\ref{fermionic}) 
as basic components for $N = 1$ SUSY and straightforwardly derive LSUSY transformations of the component fields, 
which correspond to ones for the $N = 1$ vector supermultiplet \cite{WZ,WB} with general auxiliary fields, 
by evaluating the variations (\ref{variation1}) and (\ref{variation2}) based on the commutation relation (\ref{NLSUSYcomm2}). 
In the $N = 1$ NLSUSY theory, the bosonic functionals (\ref{bosonic}) are reduced to 
\be
b = \kappa^{-1} \vert w \vert, 
\ \ b_1 = \kappa \bar\psi \psi \vert w \vert, 
\ \ b_5 = i \kappa \bar\psi \gamma_5 \psi \vert w \vert, 
\ \ b_{5a} = \kappa \bar\psi \gamma_5 \gamma_a \psi \vert w \vert, 
\ \ b_{11} = \kappa^3 \bar\psi \psi \bar\psi \psi, 
\label{bosonic1}
\ee
where we use identities which are obtained by using Fierz transformations as 
\be
\bar\psi \gamma_5 \gamma_a \psi \bar\psi \gamma_5 \gamma_b \psi 
= {1 \over 2} \eta_{ab} \left( \bar\psi \psi \bar\psi \psi - \bar\psi \gamma_5 \psi \bar\psi \gamma_5 \psi \right), 
\ \ \bar\psi \gamma_5 \psi \bar\psi \gamma_5 \psi = - \bar\psi \psi \bar\psi \psi, 
\label{b-id}
\ee
etc. The fermionic functionals (\ref{fermionic}) are also reduced to 
\be
f = \psi \vert w \vert, 
\ \ f_1 = \kappa^2 \psi \bar\psi \psi \vert w \vert, 
\label{fermionic1}
\ee
where we use identities, 
\be
\psi \bar\psi \gamma_5 \gamma_a \psi 
= -{1 \over 2} \left( \gamma_5 \gamma_a \psi \bar\psi \psi + \gamma_a \psi \bar\psi \gamma_5 \psi \right), 
\ \ \gamma_5 \psi \bar\psi \gamma_5 \psi = - \psi \bar\psi \psi, 
\label{f-id}
\ee
etc. 
By using the set of the functionals (\ref{bosonic1}) and (\ref{fermionic1}), 
we define basic component fields in $N = 1$ SUSY theories as follows; 
we express bosonic component fields as 
\be
D = b(\psi), 
\ \ A = \alpha_1 b_1(\psi), 
\ \ B = \alpha_2 b_5(\psi), 
\ \ v_a = \alpha_3 b_{5a}(\psi), 
\ \ C = \alpha_4 b_{11}(\psi), 
\label{b-comp}
\ee
and write fermionic component fields as 
\be
\lambda = f(\psi), 
\ \ \Lambda = \alpha_5 f_1(\psi), 
\label{f-comp}
\ee
in which values of constants $\alpha_m \ (m = 1,2,\cdots,5)$ are determined from definitions of fundamental actions 
in LSUSY theories and the invariances of the actions under LSUSY transformations of the components. 
Note that the "vector" field $v_a$ defined in Eq.(\ref{b-comp}) is expressed in terms of an axial-vector functional $b_{5a}$ 
and the degrees of freedom of the bosonic and fermionic components are balanced as $8 = 8$. 

Since commutation relations on the functional (composite) fields (\ref{b-comp}) and (\ref{f-comp}) 
under the NLSUSY transformation (\ref{NLSUSY}) close as Eq.(\ref{NLSUSYcomm2}) 
and the identities (\ref{b-id}) and (\ref{f-id}) are satisfied, 
the variations (\ref{variation1}) and (\ref{variation2}) for $N = 1$ SUSY 
immediately (and uniquely) determine the forms of LSUSY transformations of the basic component fields 
by using Fierz transformations: 
First, the variations of the components $D$ and $\lambda$ under NLSUSY transformations (\ref{NLSUSY}) 
are written as 
\ba
\A \A 
\delta_\zeta D = -i \bar\zeta \!\!\not\!\partial \lambda, 
\label{LSUSY-D}
\\
\A \A 
\delta_\zeta \lambda 
= D \zeta - {1 \over 4} \left( {i \over \alpha_1} \!\!\not\!\partial A 
+ {1 \over \alpha_2} \gamma_5 \!\!\not\!\partial B 
- {i \over \alpha_3} \gamma_5 \gamma^a \!\!\not\!\partial v_a \right) \zeta, 
\label{LSUSY-lambda}
\ea
for which the closure of the commutator algebra on $D$ and $\lambda$ under the LSUSY transformation's laws 
of Eqs.(\ref{LSUSY-D}) and (\ref{LSUSY-lambda}) in addition to ones of the components $(A,B,v_a)$ 
(determined below) is guaranteed by means of the commutation relation (\ref{NLSUSYcomm2}), 
though the straightforward calculations are easy, in particular, in $N = 1$ SUSY. 

Next, the variations of the components $(A,B,v_a)$ become 
\ba
\A \A 
\delta_\zeta A = \alpha_1 \left\{ 2 \bar\zeta \lambda 
- i \kappa^2 \partial_a \left( \bar\zeta \gamma^a \psi \bar\psi \psi \vert w \vert \right) \right\}, 
\label{v-A}
\\
\A \A 
\delta_\zeta B = \alpha_2 \left\{ 2i \bar\zeta \gamma_5 \lambda 
+ \kappa^2 \partial_a \left( \bar\zeta \gamma^a \psi \bar\psi \gamma_5 \psi \vert w \vert \right) \right\}, 
\\
\A \A 
\delta_\zeta v_a = \alpha_3 \left\{ 2 \bar\zeta \gamma_5 \gamma_a \lambda 
- i \kappa^2 \partial_b \left( \bar\zeta \gamma^b \psi \bar\psi \gamma_5 \gamma_a \psi \vert w \vert \right) \right\}. 
\label{v-v}
\ea
In general cases for $N \ge 2$ SUSY, deformations of $(\psi^i)^3$ terms in the second terms of those variations 
are important problems \cite{MT} in order to determine LSUSY transformations based on the commutation relations (\ref{NLSUSYcomm2}). 
However, in the $N = 1$ SUSY case, all the $\psi^3$ terms in the variations from (\ref{v-A}) to (\ref{v-v}) 
are written as $\psi \bar\psi \psi$ by means of the identities (\ref{f-id}), 
so that LSUSY transformations of $(A,B,v_a)$ are easily defined as follows; 
\ba
\A \A 
\delta_\zeta A 
= \alpha_1 \left( 2 \bar\zeta \lambda - {i \over \alpha_5} \bar\zeta \!\!\not\!\partial \Lambda \right), 
\label{LSUSY-A}
\\
\A \A 
\delta_\zeta B 
= \alpha_2 \left( 2i \bar\zeta \gamma_5 \lambda 
+ {1 \over \alpha_5} \bar\zeta \gamma_5 \!\!\not\!\partial \Lambda \right), 
\label{LSUSY-B}
\\
\A \A 
\delta_\zeta v_a 
= \alpha_3 \left( 2 \bar\zeta \gamma_5 \gamma_a \lambda 
- {i \over \alpha_5} \bar\zeta \gamma_5 \!\!\not\!\partial \gamma_a \Lambda \right), 
\label{LSUSY-v}
\ea
Furthermore, LSUSY transformations of the components $\Lambda$ and $C$ are also determined 
from the variations of their functional expressions as 
\ba
\A \A 
\delta_\zeta \Lambda 
= \alpha_5 \left\{ {1 \over 2} \left( {1 \over \alpha_1} A + {i \over \alpha_2} \gamma_5 B 
+ {1 \over \alpha_3} \gamma_5 \gamma^a v_a \right) \zeta 
- {i \over {4 \alpha_4}} \!\!\not\!\partial C \zeta \right\}, 
\label{LSUSY-Lambda}
\\
\A \A 
\delta_\zeta C = {{4 \alpha_4} \over \alpha_5} \bar\zeta \Lambda. 
\label{LSUSY-C}
\ea
Note that we use identities $\bar\psi \gamma_5 \gamma_a \psi \bar\psi \psi = \bar\psi \gamma_5 \psi \bar\psi \psi = 0$ 
in the definition of the LSUSY transformations (\ref{LSUSY-Lambda}). 

The LSUSY transformations (\ref{LSUSY-D}), (\ref{LSUSY-lambda}) and from (\ref{LSUSY-A}) to (\ref{LSUSY-C}) 
just correspond to ones for the $N = 1$ vector supermultiplet \cite{WZ,WB} 
and they satisfy the commutator algebra (\ref{NLSUSYcomm}) both on NLSUSY and LSUSY phases. 

The $U(1)$ gauge and scalar supermultiplets are obtained from appropriate recombinations of the functionals 
of the NG fermion in the above LSUSY transformations. 
As for the $U(1)$ gauge supermultiplet, we define a spinor component $\tilde \lambda(\psi)$ 
by means of the following recombination of the fermionic functionals 
such that only $U(1)$ gauge invariant quantities are produced in LSUSY transformations; namely, 
\be
\tilde \lambda(\psi) = \left( \lambda + {i \over {2 \alpha_5}} \!\!\not\!\partial \Lambda \right)(\psi). 
\label{recombi-lambda}
\ee
By using Eqs.(\ref{LSUSY-lambda}) and (\ref{LSUSY-Lambda}), 
the variation of $\tilde \lambda(\psi)$ under the NLSUSY transformation (\ref{NLSUSY}) is represented as 
\be
\delta_\zeta \tilde \lambda 
= \tilde D \zeta - {i \over {2 \alpha_3}} F_{ab} \sigma^{ab} \gamma_5 \zeta, 
\label{LSUSY-tlambda}
\ee
which is decoupled from the scalar components $(A,B)$. 
In Eq.(\ref{LSUSY-tlambda}), $F_{ab} = \partial_a v_b - \partial_b v_a$ 
and we also define an auxiliary scalar component $\tilde D(\psi)$ as 
\be
\tilde D(\psi) = \left( D + {1 \over {8 \alpha_4}} \Box C \right)(\psi). 
\label{recombi-D}
\ee
Then, the variations of $\tilde D(\psi)$ and $v_a(\psi)$ are written 
from Eqs.(\ref{LSUSY-D}), (\ref{LSUSY-v}) and (\ref{LSUSY-C}) as 
\ba
\A \A 
\delta_\zeta \tilde D = -i \bar\zeta \!\!\not\!\partial \tilde \lambda, 
\label{LSUSY-tD}
\\
\A \A 
\delta_\zeta v_a 
= 2 \alpha_3 \bar\zeta \gamma_5 \gamma_a \tilde\lambda 
+ \partial_a W, 
\label{LSUSY-tv}
\ea
where $\displaystyle{W(\psi) = -{{2i \alpha_3} \over \alpha_5} \bar\zeta \gamma_5 \Lambda(\psi)}$ means a gauge transformation parameter. 
Note that the variation (\ref{LSUSY-tv}) also satisfy the commutator algebra (\ref{NLSUSYcomm}) on LSUSY phase, 
which does not include a term for the $U(1)$ gauge transformation. This is because $W(\psi)$ satisfies a relation, 
\be
\delta_{\zeta_1} W_{\zeta_2} - \delta_{\zeta_2} W_{\zeta_1} 
= 2i \bar\zeta_1 \gamma^a \zeta_2 v_a. 
\ee
The variations (\ref{LSUSY-tlambda}), (\ref{LSUSY-tD}) and (\ref{LSUSY-tv}) constitute the LSUSY transformations 
for the $U(1)$ gauge supermultiplet in terms of the components $(\tilde \lambda, \tilde D, v_a)$. 
\footnote{
A conventional redefinition $\lambda(\psi) = i \gamma_5 f(\psi)$ may be used in the definition 
of LSUSY transformations for the $U(1)$ gauge supermultiplet. 
}

On the other hand, the scalar supermultiplet is derived by defining a spinor component $\chi(\psi)$ by means of 
the following recombination of the fermionic functionals, 
\be
\chi(\psi) = \left( \lambda - {i \over {2 \alpha_5}} \!\!\not\!\partial \Lambda \right)(\psi). 
\label{recombi-chi}
\ee
Indeed, the variation of $\chi(\psi)$ under the NLSUSY transformation (\ref{NLSUSY}) 
is given from Eqs.(\ref{LSUSY-lambda}) and (\ref{LSUSY-Lambda}) as follows; 
\be
\delta_\zeta \chi 
= \left( F + {i \over {2 \alpha_3}} G \gamma_5 \right) \zeta 
- {i \over 2} \!\!\not\!\partial \left( {1 \over \alpha_1} A + {i \over \alpha_2} B \gamma_5 \right) \zeta, 
\label{LSUSY-chi}
\ee
where auxiliary scalar components $(F,G)$ are defined as 
\be
F(\psi) = \left( D - {1 \over {8 \alpha_4}} \Box C \right)(\psi), 
\ \ G(\psi) = \partial^a v_a(\psi). 
\label{recombi-FG}
\ee
From Eqs.(\ref{LSUSY-D}), (\ref{LSUSY-A}), (\ref{LSUSY-B}), (\ref{LSUSY-v}) and (\ref{LSUSY-C}), 
the variations of $(F,G,A,B)(\psi)$ are respectively represented as 
\ba
\A \A 
\delta_\zeta F = -i \bar\zeta \!\!\not\!\partial \chi, 
\label{LSUSY-F}
\\
\A \A 
\delta_\zeta G = 2 \alpha_3 \bar\zeta \gamma_5 \!\!\not\!\partial \chi, 
\label{LSUSY-G}
\\
\A \A 
\delta_\zeta A = 2 \alpha_1 \bar\zeta \chi, 
\label{LSUSY-A2}
\\
\A \A 
\delta_\zeta B = 2 i \alpha_2 \bar\zeta \gamma_5 \chi. 
\label{LSUSY-B2}
\ea
Therefore, the variations (\ref{LSUSY-chi}) and from (\ref{LSUSY-F}) to (\ref{LSUSY-B2}) 
constitute the LSUSY transformations for the scalar supermultiplet. 

Finally, we mention LSUSY actions and the relations between the VA NLSUSY and LSUSY actions. 
A LSUSY (free) action for the $U(1)$ gauge supermultiplet is defined as 
\be
S_{U(1)\ {\rm gauge}} = \int d^4x \left\{ -{1 \over 4} (F_{ab})^2 + {i \over 2} \bar{\tilde \lambda} \!\!\not\!\partial \tilde \lambda 
+ {1 \over 2} \tilde D^2 - {1 \over \kappa} \tilde D \right\}, 
\label{gauge}
\ee
where we can take the values of $(\alpha_4, \alpha_5)$, 
e.g. as $\displaystyle{\alpha_4 = {1 \over 8}}$ and $\displaystyle{\alpha_5 = {1 \over 2}}$ 
which fix relative scales of the terms for the auxiliary fields $\Lambda$ and $C$ to the kinetic terms of the physical fields. 
The invariance of the action (\ref{gauge}) under the LSUSY transformations (\ref{LSUSY-tlambda}), (\ref{LSUSY-tD}) and (\ref{LSUSY-tv}) 
determines the value of $\alpha_3$ as $\displaystyle{\alpha_3^2 = {1 \over 4}}$. 

In addition, a LSUSY (free) action for the scalar supermultiplet is given by means of 
\be
S_{\rm scalar} = \int d^4x \left\{ {1 \over 2} (\partial_a A)^2 + {1 \over 2} (\partial_a B)^2 
+ {i \over 2} \bar\chi \!\!\not\!\partial \chi + {1 \over 2} (F^2 + G^2) - {1 \over \kappa} F \right\}. 
\label{scalar}
\ee
where we can also define the values of $(\alpha_4, \alpha_5)$, e.g. as $\displaystyle{\alpha_4 = -{1 \over 8}}$ 
and $\displaystyle{\alpha_5 = -{1 \over 2}}$. 
The invariance of the action (\ref{scalar}) under the LSUSY transformations (\ref{LSUSY-chi}) and from (\ref{LSUSY-F}) to (\ref{LSUSY-B2}) 
determines the values of $(\alpha_1, \alpha_2, \alpha_3)$ as $\displaystyle{\alpha_1^2 = \alpha_2^2 = \alpha_3^2 = {1 \over 4}}$. 

Both the LSUSY actions (\ref{gauge}) and (\ref{scalar}) are related to the VA NLSUSY action (\ref{NLSUSYaction}) 
for $N = 1$ SUSY as 
\be
S_{U(1)\ {\rm gauge}}(\psi) = S_{N = 1\ {\rm NLSUSY}}, \ \ S_{\rm scalar}(\psi) = S_{N = 1\ {\rm NLSUSY}}, 
\label{relations}
\ee
(except for surface terms) respectively, 
based on the functional (composite) relations (\ref{b-comp}) and (\ref{f-comp}) of the basic components. 
These relations (\ref{relations}) for the $N = 1$ SUSY actions have already been established 
in terms of superfields and SUSY invariant constraints in Refs.\cite{IK,Ro,STT1}. 

In this letter, we have applied a linearization procedure of NLSUSY based on the commutation relation (\ref{NLSUSYcomm2}) 
to the $N = 1$ NL/LSUSY relations. 
We have defined the basic components (\ref{b-comp}) and (\ref{f-comp}) from the set of the fermionic and bosonic 
functionals (\ref{bosonic1}) and (\ref{fermionic1}) 
and evaluated the variations (\ref{variation1}) and (\ref{variation2}) in $N = 1$ SUSY, 
so that the LSUSY transformations (\ref{LSUSY-D}), (\ref{LSUSY-lambda}) and from (\ref{LSUSY-A}) to (\ref{LSUSY-C}) 
have been constructed (uniquely), which just correspond to ones for the $N = 1$ vector supermultiplet 
with the general auxiliary fields. 
The LSUSY transformations for the $U(1)$ gauge supermultiplet have also been obtained 
from the redefinitions of basic components by means of the functional recombinations (\ref{recombi-lambda}) and (\ref{recombi-D}), 
whereas the ones for the scalar supermultiplet have been derived from the recombinations (\ref{recombi-chi}) and (\ref{recombi-FG}). 
In the relations between the VA NLSUSY action (\ref{NLSUSYaction}) for $N = 1$ SUSY 
and the LSUSY ones (\ref{gauge}) and (\ref{scalar}), 
the common NG-fermion-functional (composite) structure of the component fields is manifest, 
since the linearization procedure of NLSUSY in this letter begin with the same set of the functionals 
(\ref{bosonic1}) and (\ref{fermionic1}). 
The applications of the commutator-based linearization of NLSUSY to $N \ge 2$ SUSY theories and the NLSUSY GR one 
are interesting problems.

\vspace{7mm}

\noindent
{\large\bf Acknowledgements} \\[2mm]
The author would like to thank Kazunari Shima for valuable discussions.

\newpage

%
\newcommand{\NP}[1]{{\it Nucl.\ Phys.\ }{\bf #1}}
\newcommand{\PL}[1]{{\it Phys.\ Lett.\ }{\bf #1}}
\newcommand{\CMP}[1]{{\it Commun.\ Math.\ Phys.\ }{\bf #1}}
\newcommand{\MPL}[1]{{\it Mod.\ Phys.\ Lett.\ }{\bf #1}}
\newcommand{\IJMP}[1]{{\it Int.\ J. Mod.\ Phys.\ }{\bf #1}}
\newcommand{\PR}[1]{{\it Phys.\ Rev.\ }{\bf #1}}
\newcommand{\PRL}[1]{{\it Phys.\ Rev.\ Lett.\ }{\bf #1}}
\newcommand{\PTP}[1]{{\it Prog.\ Theor.\ Phys.\ }{\bf #1}}
\newcommand{\PTPS}[1]{{\it Prog.\ Theor.\ Phys.\ Suppl.\ }{\bf #1}}
\newcommand{\AP}[1]{{\it Ann.\ Phys.\ }{\bf #1}}

\end{document}